\documentclass{elsart}
\usepackage{graphicx}
\usepackage{amssymb}

\begin{document}

\begin{frontmatter}

\title{Enhancement of optical switching parameter and third-order optical 
nonlinearities in embedded Si nanocrystals: \\ a theoretical assessment}

\author[addlabel]{Hasan Y{\i}ld{\i}r{\i}m}
\author[addlabel]{Ceyhun Bulutay\corauthref{cor1}}
\address[addlabel]{Department of Physics, Bilkent University, Ankara 06800, Turkey}
\corauth[cor1]{E-mail: bulutay@fen.bilkent.edu.tr}

\begin{abstract}
Third-order bound-charge electronic nonlinearities of Si nanocrystals
(NCs) embedded in a wide band-gap matrix representing silica are theoretically
studied using an atomistic pseudopotential approach. Nonlinear
refractive index, two-photon absorption and optical switching
parameter are examined from small clusters to NCs up to a size of 3~nm. 
Compared to bulk values, Si NCs show higher third-order 
optical nonlinearities and much wider two-photon absorption threshold 
which gives rise to enhancement in the optical switching parameter.
\end{abstract}

\end{frontmatter}

%\section{Introduction}
Both the subjects of Si nanocrystals (NCs) and nonlinear optics are
currently very active because of their well established applications,
such as those in light emitting diodes, lasers, solar cells,
interferometers, optical switches, optical data storage elements,
and other photonic devices \cite{pavesi04}. One group of very
important optical nonlinearities is the third-order nonlinearities
which involve nonlinear refraction coefficient or
optical Kerr index $n_{2}$ and two-photon absorption coefficient
$\beta$.
These nonlinearities are crucial in all-optical switching and sensor
protection applications \cite{boyd03} as well as in the up-conversion
of the sub-band-gap light for the possible solar cell
applications \cite{thurpke}. In these respects, a clear understanding
of the third-order nonlinearities in Si NCs would play an important
role in such applications.

Si NCs with controllable sizes have electronic structure largely
affected by the quantum and dielectric confinements. Hence, they are
expected to have markedly different nonlinear optical properties
with respect to bulk Si which itself already displays significant
third-order optical nonlinearities \cite{bristow07,lin07}. In fact, recent
experimental reports
show that Si NCs have promising nonlinear optical properties and
device applications \cite{prakash02,pavesi03,king07}.
Therefore, an in-depth knowledge of Si NCs' nonlinear parameters is
essential for various nonlinear optics applications. Unfortunately, there
is neither a comprehensive theoretical work nor an experimental
study on the full wavelength and size dependence of the third-order
optical nonlinearities of Si NCs. Among the few available
experimental studies, we should mention the work of Prakash \emph{et
al.} in which $n_{2}$ and $\beta$ were measured \cite{prakash02}. 
However these measurements were performed at a single 
wavelength. For a more comprehensive understanding, a rigorous 
theoretical work is highly required that can also unambiguously extract 
the NC size effects which is not precise in the experimental studies 
due to limitations in size control in embedded NCs.

In this paper, our aim is to present such a theoretical assessment
of the third-order nonlinearities in Si NCs resolving the size
scaling and the full wavelength dependence up to UV region.
In this respect, we indiscriminately consider both nonresonant and 
resonant nonlinearities. 
Furthermore, unlike  most of the previous theoretical studies on 
linear optical properties which
considered hydrogen-passivated NCs, we deal with Si NCs embedded in
a wide band-gap matrices representing silica which is the
most common and functional choice in the actual structures \cite{pavesi04}.
The source of optical nonlinearity in this work is the bound (confined) electronic 
charge of the NC valence electrons filling up to the highest occupied molecular 
orbital. Especially for ultrafast applications in the transparency region this is the
dominant contribution \cite{bahae91}.

%\section{Theory}
The characterization of the third-order nonlinear susceptibilities
up to a photon energy of 4~eV brings a challenge for the
electronic structure. For this purpose we resort to the so-called linear
combination of bulk bands basis within the pseudopotential
framework \cite{wang}. This can handle thousands-of-atom
systems both with sufficient accuracy and efficiency over a large
energy window which becomes a major asset for the third-order
nonlinear optical susceptibilities with very demanding computational
costs of their own. Further credence for this particular computational framework 
is recently established from successful applications of quantum phenomena 
taking place over several eV energy range, such as the excited-state 
absorption \cite{bulutay07} and the Auger 
and carrier multiplication \cite{sevik08} in embedded Si and Ge NCs.
As our primary interest is on the nonlinear
optical properties, we refer to our previous work for further
details on the electronic structure \cite{bulutay07}.
However, some information regarding the embedding host matrix would be in order. 
In real applications, the common choice is silica \cite{pavesi04}. On the other hand, to 
avoid complications arising from its chemical and structural compositions, we prefer to 
replace it with an artificial medium having the same crystal structure as silicon but 
possessing the band alignment and refractive index compatible with SiO$_2$. Both of these 
are crucial for the accurate representation of the quantum and dielectric confinement 
of the actual system.

In this work, the electromagnetic interaction term in the Hamiltonian is taken
as $-e\textbf{r}\cdot\textbf{ E}$, in other words, the length gauge
is used. The third-order optical nonlinearity expressions based on
the length gauge have proved to be successful in atomic-like
systems \cite{boyd03}. Therefore, we use the expressions based on the
length gauge whose detailed forms can be found in Ref.~\cite{boyd03b}.
For embedded NCs an important parameter is the volume filling factor $f_{v}$ 
of the NCs $f_{v}=V_{NC}/V_{SC}$ where $V_{NC}$ and $V_{SC}$ are the 
volumes of the NC and the (computational) supercell, respectively. 
Throughout this work, for the sake of generality 
the nonlinear properties of the embedded NC systems are calculated 
at the unity filling factor which is indicated by  an overbar
\begin{eqnarray}\label{chi3}
\overline{\chi}^{(3)}_{dcba}(-\omega_{3};\omega_{\gamma},\omega_{\beta},\omega_{\alpha})&\equiv
&\frac{\chi^{(3)}_{dcba}(-\omega_{3};\omega_{\gamma},\omega_{\beta},\omega_{\alpha})}{f_{v}}
\end{eqnarray}
where the subscripts $\{a,b,c,d\}$ refer to cartesian indices,
$\omega_{\gamma}$, $\omega_{\beta}$, and $\omega_{\alpha}$ are the
input frequencies,
$\omega_{3}\equiv\omega_{\gamma}+\omega_{\beta}+\omega_{\alpha}$.
Our results can trivially be converted to any specific filling factor.
As we consider only spherical NCs, we set the cartesian tensor indices as 
$\{a,b,c,d\}$=$\{1,1,1,1\}$ and suppress these subscripts for convenience. 
Furthermore, we use a Lorentzian energy broadening parameter of 100~meV at 
full width. The value refers to the typical room temperature photoluminescence (hence, interband) 
linewidth of a \emph{single} Si NC  embedded in SiO$_2$ as considered in this work \cite{valenta02}.
This parameter not only broadens the resonances but also introduces the 
band tailing effects \cite{lloyd} which becomes especially important in 
the transparency regions.

At high illumination intensities, third-order changes  in the refractive 
index and the absorption are observed due to the virtual and real excitations 
of the bound charges. Accounting for these effects, the refractive index and the
absorption become, respectively, $n=n_{0}+n_{2}I$ and
$\alpha=\alpha_{0}+\beta I$, where $n_{0}$ is the linear refractive
index, $\alpha_{0}$ is the linear absorption coefficient, and
\textit{I} is the light intensity. $\overline{n}_{2}$ and
$\overline{\beta}$ are proportional to
$\textrm{Re}\overline{\chi}^{(3)}(-\omega;\omega,-\omega,\omega)$
and $\textrm{Im}\overline{\chi}^{(3)}(-\omega;\omega,-\omega,\omega)$,
respectively \cite{bahae90}. Furthermore, the degenerate 
two-photon absorption cross section \cite{boyd03} at
unity filling factor is given by
$\overline{\sigma}^{(2)}(\omega)\equiv\sigma^{(2)}(\omega)/f_{v}$;
$\overline{\sigma}^{(2)}(\omega)$ and $\overline{\beta}$ are related
to each other through
$\overline{\beta}=2\hbar\omega\overline{\sigma}^{(2)}(\omega)$.

In the case of NCs, one should take into account the local field effects 
(LFEs) as in
composite materials the dielectric mismatch between the constituents may lead to remarkably different optical
properties \cite{sipe02}.
Incorporation of the LFEs into calculations is not a trivial task.
For structures of the so-called Maxwell-Garnett
geometry \cite{sipe02}, the LFEs yield the following correction
factor for the third-order nonlinear optical properties \cite{sipe02}
\begin{equation}\label{lfe}
L=\left(\frac{3\epsilon_{h}}{\epsilon_{NC}+2\epsilon_{h}}\right)^{2}
\left|\frac{3\epsilon_{h}}{\epsilon_{NC}+2\epsilon_{h}}\right|^{2}
\end{equation}
where $\epsilon_{h}$ and $\epsilon_{NC}$ are  the dielectric
functions of the host matrix and the NC, respectively. Note that in
this equation it is assumed that the inclusions are
non-interacting spheres and the host matrix does not show any
significant nonlinearity \cite{sipe92}. 
In our implementation we use a static local field correction, otherwise
the correction factor spuriously causes negative absorption regions at high energies.
Further theoretical work is much needed in this direction.
Regarding the other shortcomings of our model, we would like to mention 
that our treatment does not include the excitonic, strain, thermal and 
free-carrier effects, which may form the basis of possible extensions 
of this work. Nevertheless, we can argue that in our context the lack of these effects will not have
qualitative consequences. First of all regarding the strain, a very recent and realistic 
atomistic modeling of this system has revealed that the inner core of the NCs where 
most wave function localization occurs remains unstrained \cite{yilmaz08}. 
As for the thermal effects, in this work they are indirectly accounted through the Lorentzian 
broadening parameter which is predominantly caused by the low energy acoustic 
phonons \cite{sychugov05}. On the other hand, the free-carrier, phonon-assisted or other thermal nonlinearities 
are totally left out of the scope of our treatment which focuses on the bound-charge nonlinearities 
taking place in much faster time scale. Finally, the excitonic effects may introduce new features 
to the spectra, however, at room temperature we do not expect them to markedly stand out.

\begin{figure}
\includegraphics[width=13 cm, angle=0]{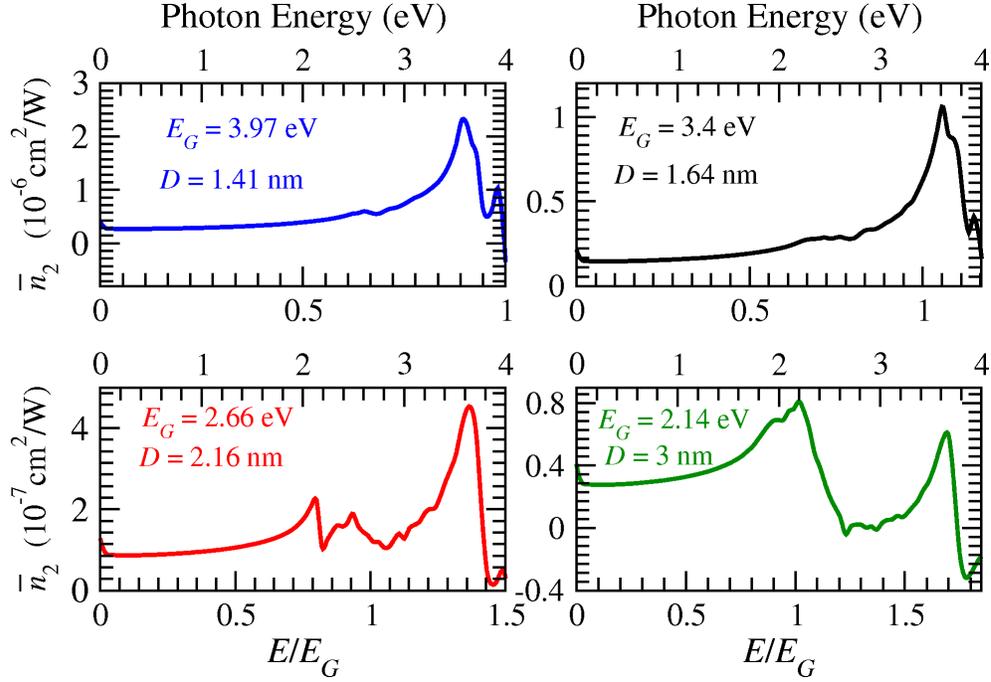}
\caption{\label{fig-1}
$\overline{n}_{2}$ as a function of the photon energy for different NC sizes.}
\end{figure}

%\section{Results and Discussion}
We consider four different diameters, $D=$1.41, 1.64, 2.16 and 3~nm.
Their energy gap $E_G$ as determined by the separation between the lowest 
unoccupied molecular orbital (LUMO) and the highest occupied molecular orbital 
(HOMO) energies
show the expected quantum size effect \cite{bulutay07}.
The $\overline{n}_{2}$ is  plotted in Fig.~\ref{fig-1} which increases with 
the decreasing NC size for all
frequencies. The smallest diameter gives us the largest
$\overline{n}_{2}$. When compared to the $n_{2}$ of bulk Si in this
energy interval
($\sim10^{-14}$~cm$^{2}$/W) \cite{bristow07,lin07,dinu03}, our
calculated $n_{2}$ is enhanced as much as ($\sim10^{6}f_{v}$) for
the largest NC. For Si NCs having a diameter of a few nanometers,
Prakash \emph{et al.} \cite{prakash02} have obtained $n_{2}$ of the
order of $\sim10^{-11}$~cm$^{2}$/W which in order of magnitude agrees
with our results when a typical $f_{v}$ is assumed for their samples.

\begin{figure}
\includegraphics[width=13 cm, angle=0]{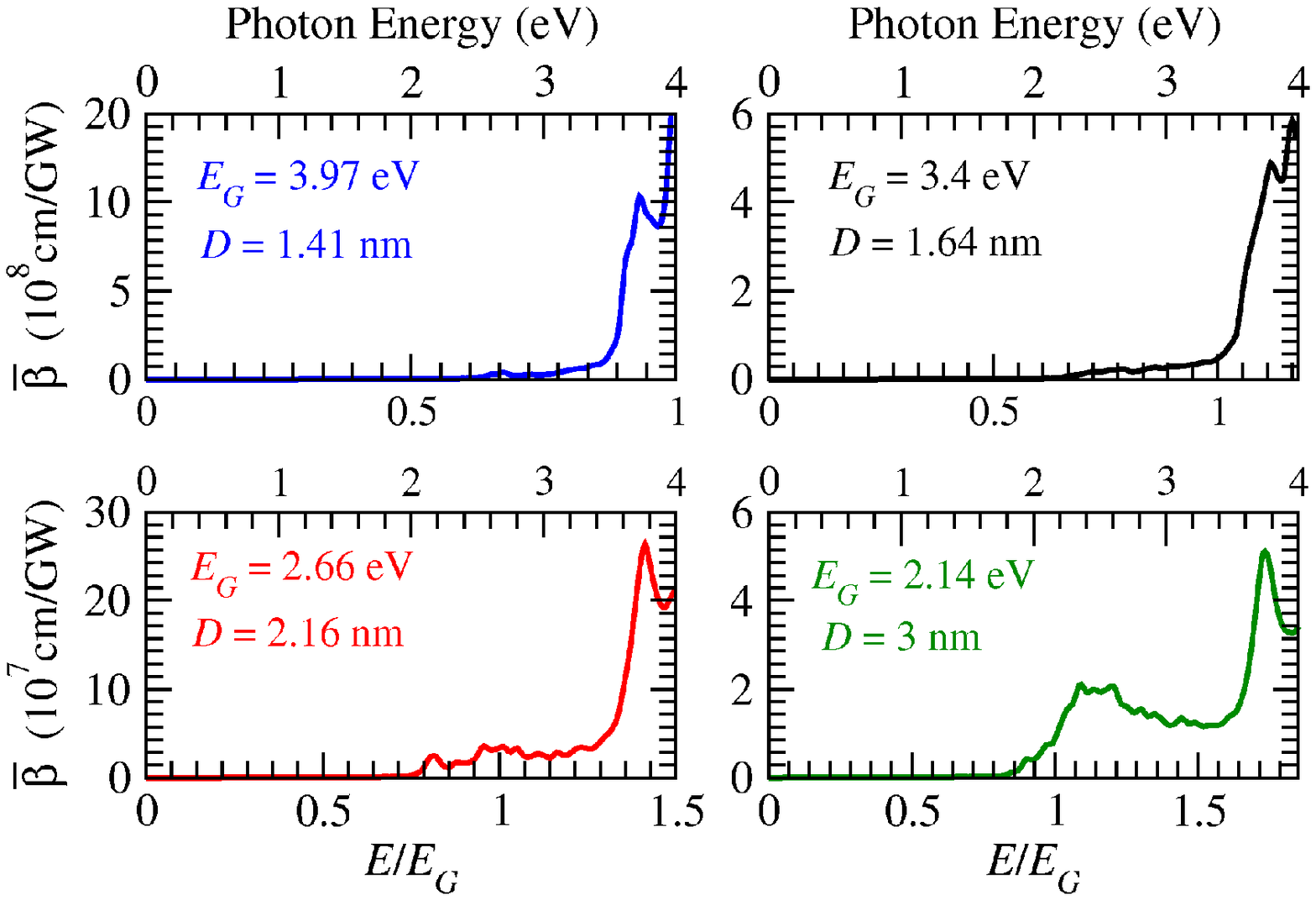}
\caption{\label{fig-2}
$\overline{\beta}$ as a function of the photon energy for different NC sizes.}
\end{figure}

\begin{figure}
\includegraphics[width=13 cm, angle=0]{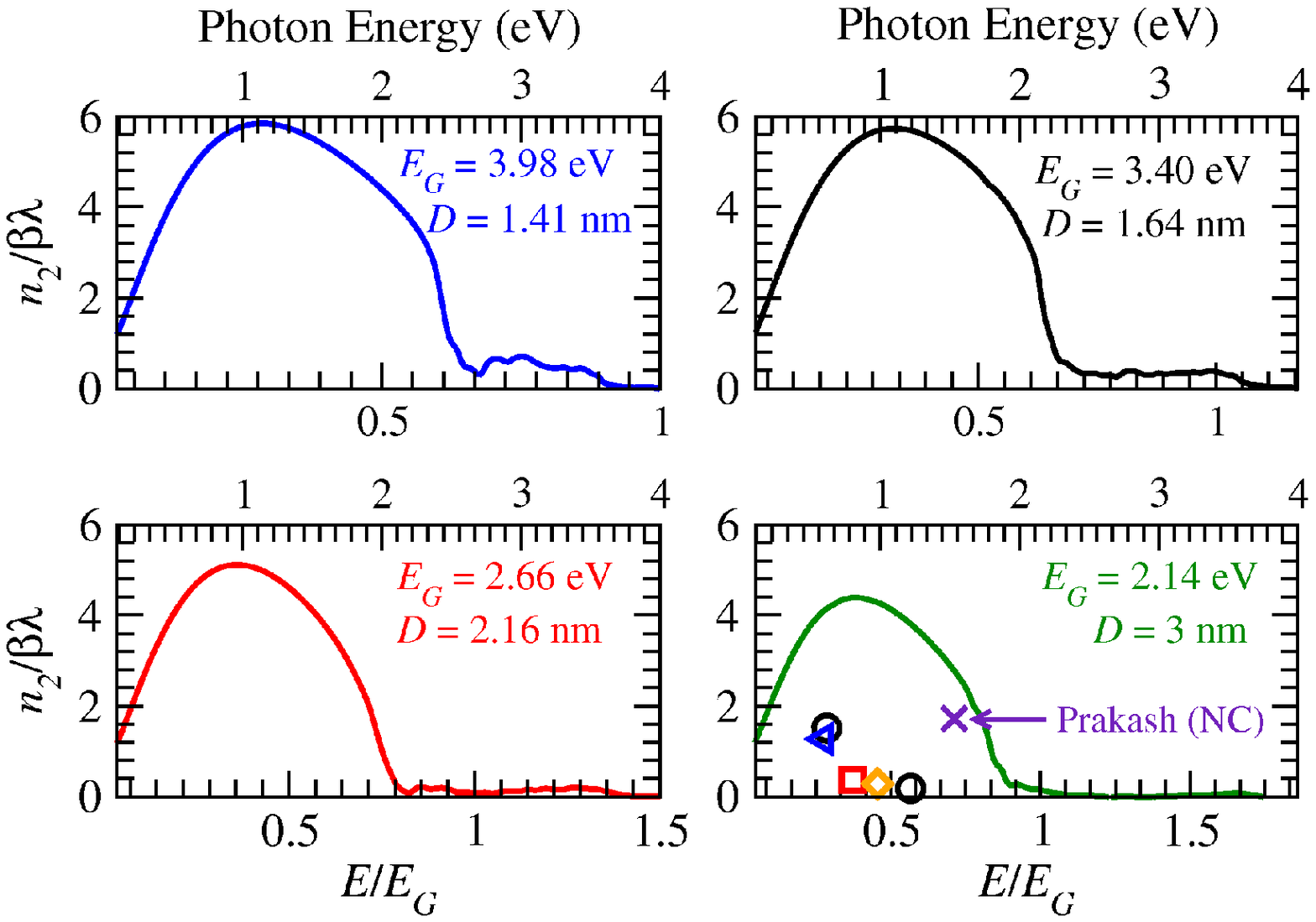}
\caption{\label{fig-3}
Optical switching parameter $n_{2}/\beta\lambda$ as a function of the photon energy for different NC sizes.
The symbols refer to experimental values for bulk Si indicated by diamond \cite{dinu03}, square \cite{dinu03}, 
circle \cite{bristow07} and triangle \cite{lin07} and NC Si indicated by cross \cite{prakash02}.}
\end{figure}

We have plotted $\overline{\beta}$ against the
photon energy in Fig.~\ref{fig-2}. Peaks at high energies are
dominant in the spectrum and $\overline{\beta}$ decreases with the
growing NC volume.
The obtained $\beta$ is about
$10^{5}f_{v}$~cm/GW for the largest NC at around 1~eV.  When compared to the
experimental bulk value ($1.5-2.0$~cm/GW measured at around 1~eV) \cite{bristow07}, our
calculated $\beta$ is enhanced about $10^{5}f_{v}$ times. Prakash
\emph{et al} \cite{prakash02} have observed $\beta$ to be between
($10^{1}-10^2$~cm/GW) at $1.53$ eV which is close to our values
provided that $f_{v}$ is taken into account. 
Moreover, $|\chi^{(3)}(-\omega;\omega,-\omega,\omega)|$ at
$1.53$~eV in our results (not shown) is enhanced with the decreasing
volume. This is in full agreement with the findings of Prakash \emph{et al} 
\cite{prakash02}. However, the result of the composite material was given in this experiment rather 
than the contributions of each constituent, that is,  the host matrix and Si NCs. 

We should note that $\overline{\beta}$ is nonzero down to 
static values due to band tailing as mentioned above. 
Another interesting observation is that the two-photon absorption 
threshold is distinctly beyond the half band gap value, which becomes more 
prominent as the NC size increases. 
This can be explained mainly as the legacy of the NC core medium, silicon which is an indirect band-gap 
semiconductor. Hence, the HOMO-LUMO dipole transition is very weak especially for relatively large 
NCs. We think that this is the essence of what is observed also for 
the two-photon absorption. As the NC size gets smaller, the HOMO-LUMO energy gap 
approaches to the direct band-gap of bulk silicon, while the HOMO-LUMO dipole transition 
becomes more effective.

In optical switching systems $n_{2}$ has an important
role \cite{boyd03}. A good optical switching device should possess
the condition $\Delta n>c_{sw}\alpha\lambda$ where $\Delta n$ is the
change in the refractive index and $c_{sw}$ is a constant of the
order of unity, the exact value of which depends on the switching
system \cite{bahae91}. For the energies below the band gap, one has
$\Delta \alpha=\beta I$ and $\Delta n=n_{2}I$. In this case, the condition
reduces to $n_{2}/\beta\lambda>c_{sw}$. We have plotted this 
optical swicthing parameter
$n_{2}/\beta\lambda$ as a function of the photon energy in Fig.~\ref{fig-3}.
Their general behavior resembles the results of Khurgin and Li for the
third-order intersubband nonliearities in quantum wells \cite{khurgin93}.

The condition gives us values  already exceeding unity and it brings
about an immediate peak at around $1$~eV  for each diameter. 
We should note that this is not the case for bulk Si which shows a 
monotonic behavior, as computed by Dinu for the phonon-assisted 
two-photon absorption \cite{dinu03b}. 
It can be also observed from this figure that the peak positions are red-shifted 
with the increasing NC volume.
Notably, the switching condition gives higher values as the NC gets
smaller, but it converges to a specific value and does not get enhanced
further. 
In the same figure some experimental data for the bulk
Si \cite{bristow07,lin07,dinu03} and Si NC with $D=3$~nm \cite{prakash02}
are also shown. Note that the experimental result for Si
NCs \cite{prakash02} is higher than the bulk values. Our result for
the NC with $D=3$ nm is in  good agreement with the experimental NC
value. Our calculations lie even above the experimental ones.
However, we think that a better agreement will be obtained when the
experiments are held at a broad range of laser wavelengths. As a
result, the calculations show that Si NCs access large values of the
ratio $n_{2}/\beta\lambda$, particularly at small NC volumes. This feature 
should be taken into account in assessing Si NCs for nonlinear device
applications especially for optical switching systems \cite{bahae91,khurgin93}.

%\section{Conclusion}
In summary, we have investigated the wavelength and size dependence
of the bound-charge third-order optical nonlinearities in Si NCs. It is observed
that  both $n_{2}$ and $\beta$ are enhanced with the decreasing of
the NC size, giving values greater than their respective bulk
values. Finally, optical switching parameter is assessed based on
the numerical results. Si NCs enhance this parameter with respect to
bulk Si.

\textbf{Acknowledgments}\\
This work has been supported by the European Commission's FP6 Project SEMINANO
under Contract No. NMP4-CT2004-505285 and by the Scientific and Techological Research 
Council of Turkey, T\"{U}B\.ITAK under the B\.IDEB Programme and 
with the Project No. 106T048.

% References in Opt. Comm. format


\begin{thebibliography}{99}
\bibitem{pavesi04}L. Pavesi and D.J. Lockwood, Silicon Photonics, Springer, Berlin, 2004.
\bibitem{boyd03} R.W. Boyd, Nonlinear Optics, Academic Press, San Diego, 2003.
\bibitem{thurpke}T. Thurpke, M.A. Green, and P. W\"urfel, J. Appl. Phys. 92 (2002) 4117.
\bibitem{bristow07} A.D. Bristow, N. Rotenberg, and H.M. van Driel, Appl. Phys. Lett. 90 (2007) 191104.
\bibitem{lin07}Q. Lin, J. Zhang, G. Piredda, R.W. Boyd, P.M. Fauchet, and G.P. Agrawal, Appl. Phys. Lett. 91 (2007) 021111.
\bibitem{prakash02} G.V. Prakash, M. Cazzanelli, Z. Gaburro, L. Pavesi, F. Iacona, F. Franzo, and J.G. Priolo, J. Appl. Phys. 91 (2002) 4607.
\bibitem{pavesi03}L. Pavesi, Z. Gaburro, L. Dal Negro, P. Bettotti, G.V. Prakash, M. Cazzanelli, and C.J. Oton, Optic Las. Eng. 39 (2003) 345. 
\bibitem{king07}S.M. King, S. Chaure, J. Doyle, A. Colli, A.C. Ferrari, and W.J. Blau, Opt. Commun. 276  (2007) 305. 
\bibitem{bahae91} M. Sheik-Bahae, D.C. Hutchings, D.J. Hagan, and E.W. Van Stryland, IEEE J. Quantum Electron. 27 (1991) 1296.
\bibitem{wang} L.W. Wang, A. Franceschetti, and A. Zunger, Phys. Rev. Lett. 78 (1997) 2819.
\bibitem{bulutay07} C. Bulutay, Phys. Rev. B 76 (2007) 205321.
\bibitem{sevik08} C. Sevik and C. Bulutay, Phys. Rev. B 77 (2008) 125414.
\bibitem{boyd03b} R.W. Boyd, Nonlinear Optics, p. 173, Academic Press, San Diego, 2003.
\bibitem{valenta02} J. Valenta, R. Juhasz, and J. Linros, Appl. Phys. Lett. 80 (2002) 1070.
\bibitem{lloyd}P. Lloyd, J. Phys. C: Solid State Phys. 2 (1969) 1717.
\bibitem{bahae90} M. Sheik-Bahae, A.A. Said, T.H. Wei,  D.J. Hagan, and E.W. Van Styrland, IEEE J. Quantum Electron. 26 (1990) 760.
\bibitem{sipe02} J.E. Sipe and R.W. Boyd, Optical Properties of Nanostructured Random Media, in: V. M. Shalev (Ed.), Topics
Appl. Phys., vol. 82, Springer, Berlin-Heidelberg, 2002.
\bibitem{sipe92} J.E. Sipe and R. W. Boyd, Phys. Rev. A 46 (1992) 1614.
\bibitem{yilmaz08} D. E. Y{\i}lmaz, C. Bulutay, and T. \c{C}a\u{g}{\i}n, Phys. Rev. B 77 (2008) 155306.
\bibitem{sychugov05} I. Sychugov, R. Juhasz, J. Valenta, and J. Linros, Phys. Rev. Lett. 94 (2005) 087405.
\bibitem{dinu03} M. Dinu, F. Quochi, and H. Garcia, Appl. Phys. Lett. 82 (2003) 2954.
\bibitem{khurgin93}  J.B. Khurgin and S. Li, Appl. Phys. Lett. 62 (1993) 126.
\bibitem{dinu03b} M. Dinu, IEEE J. Quantum Electron. 39 (2003) 1498.
\end{thebibliography}
\end{document}